\documentclass[a4paper,nofootinbib,10pt,showpacs,showkeys]{revtex4}

\usepackage{amssymb}
\usepackage{amsmath}
\usepackage{graphics}
\usepackage{graphicx}
\usepackage{dcolumn}
\usepackage{bm}
\usepackage{color}
\usepackage{hyperref} 

\begin{document}
	
	\newcommand{\blue}[1]{\textcolor{blue}{#1}}
	\newcommand{\new}{\blue}
	
	\baselineskip=15pt

	
	%
	%
	
\title{The Debye Length and the Running Coupling of QCD: a Potential and Phenomenological Approach}
	
\author{S. D. Campos} \email{sergiodc@ufscar.br}
\affiliation{Applied Mathematics Laboratory-CCTS/DFQM, Federal University of \\ S\~ao Carlos, Sorocaba, S\~ao Paulo CEP 18052780, Brazil}

	
\begin{abstract}
		In this paper, one uses a damped potential to present a description of the running coupling constant of QCD in the confinement phase. Based on a phenomenological perspective for the Debye screening length, one compares the running coupling obtained here with both the Brodsky-de T\'eramond-Deur and the Richardson approaches. The results seem to indicate the model introduced here corroborate the Richardson approach. Moreover, the Debye screening mass in the confinement phase depends on a small parameter, which tends to vanish in the non-confinement phase of QCD.
\end{abstract}
	


\maketitle

\section{Introduction}\label{sec:intro}

The understanding of non-perturbative dynamics of Quantum Chromodynamics (QCD), from first principles, is a major goal of high energy physics. Although less fundamental than lattice QCD or QCD sum rules, models based on the interaction potential approach has been proven to be very useful, even in the non-relativistic approximation. As well-known, quarks and gluons in the hadron can be regarded as relativistic or non-relativistic, depending on the strength of the interaction potential. For example, the Cornell confinement potential \cite{e_eichten_Phys_Rev_Lett_34_369_1975,eichten_Phys.Rev.D17.3090.1978,e_eichten_Phys.Rev.D21.203.1980} can be used to study a non-relativistic system of heavy hadrons, whereas the potential of De R\'ujula \textit{et al.} \cite{A.De.Rujula.H.Georgi.S.L.Glashow.Phys.Rev.D.12.147.1975} is applied to describe light quarks \cite{R.K.Bhaduri.L.E.Cohler.Y.Nogami.Phys.Rev.Lett.44.1369.1980}. There is no surprise in the fact that some non-relativistic potentials can describe mesons composed of heavy quarks as well as some relativistic $s\bar{s}$ states \cite{a.martin.phys.lett.b93.338.1980,a.martin.phys.lett.b100.511.1988,a.martin.phys.lett.b21.561.1980}. The relativistic/non-relativistic problem is addressed to \cite{w.lucha.phys.rev.d46.1088.1992,g.jaczko.phys.rev.d58.114017.1998}, while the ways of mathematically confine something can be found in \cite{e_eichten_Phys_Rev_Lett_34_369_1975,eichten_Phys.Rev.D17.3090.1978,e_eichten_Phys.Rev.D21.203.1980,a.martin.phys.lett.b93.338.1980,a.martin.phys.lett.b100.511.1988,a.martin.phys.lett.b21.561.1980,C.Quigg.J.L.Rosner.Phys.Lett.B71.153.1977,C.Quigg.J.L.Rosner.Phys.Rep.56.4.167.1979,J.L.Richardson.Phys.Lett.B82.272.1979,x.song.z.phys.c34.223.1987,d.b.lichtenberg.z.phys.c41.615.1989}, each model being useful in some ground. 

The base of the finite-temperature QCD is the running coupling constant, $\alpha_s$, assumed to be a temperature-dependent function. At low-temperatures, one has the gas of hadronic resonances as a manifestation of the quarks and gluons in the so-called confinement phase. In general, when the temperature (or density) in the system is high enough, the hadronic matter undergoes a phase transition transforming itself in the so-called Quark-Gluon Plasma (QGP). The fundamental symmetries are, then, evidenced by the presence of free quarks and gluons with the consequent restoration of the chiral symmetry at an energy density $\sim$ 1.0 GeV/fm$^3$, forming the charmonium \cite{M.A.Shifman.A.I.Vainshtein.V.I.Zakharov.Nucl.Phys.B.147(5).448.1979}. There is experimental evidence for the formation of QGP since the fundamental results of SPS (CERN) \cite{J.P.Baizot.Nulc.Phys.A661.3c.1999,E.V.Shuryak.Nucl.Phys.A661.119c.1999} and RHIC (BNL) \cite{I.Arsene.etal.BRAHMS.Coll.Nucl.Phys.A757.1.2005,K.Adcox.etal.PHENIX.Coll.Nucl.Phys.A757.184.2005,B.B.Back.etal.PHOBOS.Coll.Phys.Rev.C72.051901.2005,J.Adams.etal.STAR.Coll.Nucl.Phys.A757.102.2005}. The identification of QGP is performed indirectly \textit{viz.} by the observation of suppression of $J/\Psi$ particle, jet quenching, enhanced production of strange particles, excess production of photons and dileptons. There is a vast literature about these issues (see, for example, \cite{H.Satz.Lect.Notes.Phys.945.1.2018} and references therein).

From the lattice QCD viewpoint, supposing a vanishing chemical potential, then the transition from confinement to non-confinement phase occurs at the temperature $T\approx 170\sim 190$ MeV \cite{M.Cheng.etal.Phys.Rev.D74.054507.2006,Y.Aoki.etal.JHEP.0906.088.2009}. In contrast, for a non-vanishing chemical potential, the predictions differ since different lattice quantities can be used \cite{Z.Fodor.S.D.Katz.Phys.Lett.B534.87.2002,Z.Fodor.S.D.Katz.K.K.Szabo.Phys.Lett.B568.73.2003,M.Kitazawa.T.Koide.T.Kunihiro.Y.Nemoto.Phys.Rev.D65.091504.2002} (see \cite{S.Borsanyi.etal.arXiv.2002.02821v1.heplat.2020} and references therein).   

In the present work, one uses a damped version of the Cornell confinement potential to study both the Debye screening length, $\lambda$, a fundamental quantity describing the hadron-quark transition at high-temperature \cite{T.Tatsumi.D.N.Voskresensky.OMEG.03.2003}, and the running coupling constant of QCD. As well-known, the Debye length in the non-confinement phase (that occurs at the ultra-violet domain - UV) depends on the running coupling constant $\alpha_s$, which defines the inverse of the Debye length, known as the Debye screening mass. Here, based on some physical assumptions, one assumes the Debye length in the confinement phase (that occurs at the infra-red domain - IR) follows the same general behavior presented by the decreasing of the hadronic total cross section, which occurs from Coulomb-Nuclear interference (CNI) up to its minimum. Thus, from a simple toy-model, one can fit the experimental data for the proton-proton ($pp$) total cross section, which will furnish an indicative of the Debye length behavior in the IR domain. Unfortunately, there is no agreement on the definition of $\alpha_s$ in the confinement phase \cite{A.Deur.S.J.Brodsky.G.F.deTeramond.Prog.Part.Nucl.Phys.90.1.2016}. Thus, one compares the running coupling in the confinement phase obtained here with the approaches of Brodsky-de T\'eramond-Deur (BdTD) \cite{S.J.Brodsky.G.F.deTeramond.A.Deur.Phys.Rev.D81.096010.2010} and the Richardson \cite{J.L.Richardson.Phys.Lett.B82.272.1979}. Comparisons with lattice QCD results are also performed \cite{O.Kaczmarek.F.Karsch.P.Petreczky.F.Zantow.Phys.Rev.D70.074505.2004.Erratum.ibid.D72.059903.2005,O.Kaczmarek.F.Zantow.Phys.Rev.D71.114510.2005,O.Kaczmarek.PoS.2008}.

The paper is organized as follows. In Section \ref{sec:damped}, one presents a simple damped confinement potential, obtaining the hadron radius. As a simple example, one extracts the proton radius, obtaining $\approx 0.83$ fm, in agreement with recent measurement \cite{W.Xiong.etal.Nature.575.147.2019}. In Section \ref{sec:tcs}, one uses a simple toy-model to fit the $pp$ experimental data for the total cross section. In Section \ref{sec:debye}, one proposes the Debye length follows the same phenomenological trend given by the decreasing of the total cross section of the toy-model introduced in the preceding section. Also, one introduces a novel formulation for the damped potential. The running coupling constant obtained is, then, confronted with the approaches due to BdTD and the Richardson. Finally, a summary and discussion of the results are presented in Section \ref{sec:critical}.

\section{Damped Confinement Potential}\label{sec:damped}

There are several formulations for the confinement potential adopted in QCD \cite{a.martin.phys.lett.b93.338.1980,a.martin.phys.lett.b100.511.1988,a.martin.phys.lett.b21.561.1980,x.song.z.phys.c34.223.1987,d.b.lichtenberg.z.phys.c41.615.1989}. The first version of the Cornell potential can be written as \cite{e_eichten_Phys_Rev_Lett_34_369_1975,eichten_Phys.Rev.D17.3090.1978,e_eichten_Phys.Rev.D21.203.1980}
\begin{eqnarray}
V(r)=-\frac{\eta}{r}+\frac{r}{a^2}+C,\label{cornell_1}
\end{eqnarray}

\noindent where $r$ is the distance between the infinitely heavy static quark-antiquark pair ($q\bar{q}$-pair) and $a^{-2}$ is the string tension. The constants $\eta\approx 0.48$, $a\approx 2.34$ GeV$^{-1}$ and $C=-0.25$ GeV were obtained originally from the charmonium spectrum \cite{eichten_Phys.Rev.D17.3090.1978,e_eichten_Phys.Rev.D21.203.1980}. In a modern view, the above potential depends on the definition of the so-called running coupling constant of QCD, written in one-loop approximation as \cite{PDG-PhysRev-D98-030001-2018}
\begin{eqnarray}\label{eq:def_alpha}
\alpha_s(\mu)=\frac{1}{4\pi\beta_{0}\ln\bigl(\mu^{2}/\Lambda_{\scriptsize{\mbox{QCD}}}^{2}\bigr)},
\end{eqnarray}

\noindent being responsible by the strong interaction at a specific energy scale $\mu$, and where $\beta_{0} = (33-2n_{f})/12\pi^2$ is written in terms of the $1$-loop $\beta$-function \cite{PDG-PhysRev-D98-030001-2018}. The number of active quark flavors at the energy scale $\mu$ is given by $n_{f}$: $n_f=6$ for $\mu\geq m_t$, $n_f=5$ for $m_b\leq \mu \leq m_t$, $n_f=4$ for $m_c\leq \mu \leq m_b$ and $n_f=3$ for $\mu\leq m_c$ \cite{buras_book}. The $\Lambda_{\scriptsize{\mbox{QCD}}}$-parameter depends on both the renormalization scheme and the flavor number $n_{f}$ \cite{PDG-PhysRev-D98-030001-2018}.

In general, one can identify $\mu=Q$, where $Q$ is the transferred momentum in the gluon frame (the probe scale). The relativistic regime demands large $Q$ (small distances in the UV domain). Furthermore, the existence of the confinement phase is related to the growth of $\alpha_s(Q)$ for small $Q$ (large distances in the IR domain, non-relativistic regime).

In general, the sine-Fourier transform of $\alpha_{s}(Q)$ can be performed (analytically or numerically) to write $\alpha_s(r)$ \cite{D.V.Shirkov.Theor.Math.Phys.136.1.893.2003}. The resulting Cornell potential can be written as
\begin{eqnarray}\label{eq:model_2}
V(r)=-\frac{4}{3}\frac{\alpha_s(r)}{r}+\sigma r.
\end{eqnarray} 

On one hand, for small distances (UV domain), the potential (\ref{eq:model_2}) has the expected Coulomb-like behavior, and the running coupling $\alpha_s(r)$ measures the interaction strength, playing a similar role to the electron charge in electrodynamics. On the other hand, for large distances (IR domain), the coupling strength is given by the string tension $\sigma=a^{-2}$  \cite{PRD-90-074017-2014}, and the potential (\ref{eq:model_2}) exhibit the experimental feature, which means the pairs behave as a color singlet (the asymptotic freedom). Notice the string tension value may vary according to the approach and, for example, the analyses of charmonium result in $\sqrt{\sigma}=394(7)$ MeV \cite{T.Kawanai.S.Sasaki.Phys.Rev.Lett.107.091601.2011,T.Kawanai.S.Sasaki.Phys.Rev.D85.091503.2012} whereas, for static sources, one obtains $\sqrt{\sigma}=460$ MeV \cite{S.Aoki.etal.PACS-CS.Collaboration.Phys.Rev.D79.034503.2009,Y.Koma.M.Koma.Proc.Sci.LATTICE2012.140.2014.arXiv:1211.6795}. In this work, without loss of generality, one defines $\sqrt{\sigma}=400$ MeV.

The phase transition from the confinement to the non-confinement occurs at $r_{\tiny{\mbox{0}}}$, the zero of the potential (\ref{eq:model_2}), which depends on the ratio of the running coupling constant to string tension
\begin{eqnarray}\label{eq:model_2.1}
r_{\tiny{\mbox{0}}}=2\sqrt{\frac{\alpha_s(r_{\tiny{\mbox{0}}})}{3\sigma}}.
\end{eqnarray}

This root corresponds to the spatial scale for the phase transition in the sense that for $r<r_{\tiny{\mbox{0}}}$, the $q\bar{q}$-pairs are subject to a Coulomb-like potential, behaving as a free (color) charged gas. Considering $r>r_{\tiny{\mbox{0}}}$, then they behave as the singlet, mentioned above. Thus, there is a transition from a non-charged ($r>r_{\tiny{\mbox{0}}}$) to a charged gas ($r<r_{\tiny{\mbox{0}}}$). Although the value of the confinement scale $r_{\tiny{\mbox{0}}}$ is not accessible to experiments, its value can be estimated from the heavy quarkonium phenomenology. From the bottomonium lattice calculation, one has $r_{\tiny{\mbox{0}}}\approx 0.47$ fm \cite{A.Gray.etal.Phys.Rev.D72.094507.2005}. Here, one uses the approach developed in \cite{C.Quigg.J.L.Rosner.Phys.Lett.B71.153.1977,C.Quigg.J.L.Rosner.Phys.Rep.56.4.167.1979}, defining
\begin{eqnarray}\label{eq:rc}
r_{q\bar{q}}=\frac{1}{\mu(q\bar{q})},
\end{eqnarray}
	
\noindent where $\mu(q\bar{q})$ is the reduced mass of the different $q\bar{q}$-pairs: $u\bar{u}$ or $d\bar{d}$. Then, the confinement scale is given by $r_{\tiny{\mbox{0}}}=\sqrt{r_{u\bar{u}}r_{d\bar{d}}}\approx 0.6$ fm.

In a more general picture, however, the order parameter concerning the confinement/non-confinement phase transition depends on some critical temperature $T_{\tiny{\mbox{L}}}$, corresponding to the Polyakov loop $P_{\tiny{\mbox{L}}}$: when $T<T_{\tiny{\mbox{L}}}$, the symmetry gauge group is unbroken and $P_{\tiny{\mbox{L}}}= 0$; when $T>T_{\tiny{\mbox{L}}}$, the symmetry is spontaneously broken  and $P_{\tiny{\mbox{L}}}\neq 0$ \cite{a.m.polyakov.nucl.phys.b120.429.1977}. It is also interesting to note that the confinement/non-confinement phase transition is an example of an inverse-melting type process \cite{n.avraham.nature.411.451.2001,a.l.gree.j.l-c.metals.140.327.1988}.

Notice that for the constituents of the pair separated by the largest distance allowable in the system, then one has a measure of the maximum of the interaction strength as well as the effective hadron size. The finiteness of the hadron is a consequence of the increasing $\alpha_s(Q)$, that should be stopped at the infrared scale since the wavelength associated with the particle created cannot exceed the hadron size \cite{brodsky_phys_lett_b666_95_2008}. 

Taking into account the above discussion, the finite size of the hadron leads to the use of (non-relativistic) screened damped confinement potentials \cite{T.Matsui.H.Satz.Phys.Lett.B.4.178.416.1986,K.Kanaya.H.Satz.Phys.Rev.D.34.10.3193.1986,F.Karsch.M.T.Mehr.H.Satz.Z.Phys.C.37.617.1988}, whose main goal is the study the possible deconfinement of heavy quarks at some transition temperature. An example of such potential is given in \cite{F.Karsch.M.T.Mehr.H.Satz.Z.Phys.C.37.617.1988} 
\begin{eqnarray}\label{eq:karsch_1}
V(r,T)=\frac{\sigma}{\mu(T)}(1-e^{-\mu(T)r})-\frac{\alpha_{\tiny{\mbox{eff}}}}{r}e^{-\mu(T)r},
\end{eqnarray}

\noindent where $\mu(T)=1/\lambda(T)$ is the inverse of the Debye screening length $\lambda(T)$, and the parameter $\alpha_{\tiny{\mbox{eff}}}$ stands for the {\it effective} running coupling. Both parameters depend on the physical properties of the plasma as well as the system temperature. It should be stressed that the temperature is an important key to elucidate the possible dissociation from hadronic to quark matter since hadrons are considered as bound states of quarks. In some lattice QCD calculations, the ab initio calculation of the thermodynamic properties of the hadron allows the extraction of the transition temperature \cite{H.Satz.Cern.TH.7410.94,O.Kaczmarek.F.Zantow.Phys.Rev.D71.114510.2005,O.Kaczmarek.etal.Phys.Rev.D83.014504.2011,C.Schmidt.J.Phys.Conf.Ser.432.012013.2013,P.W.M.Evans.C.R.Allton.J.I.Skullerud.Phys.Rev.D89.071502.2014}. For a recent status of hot-dense lattice QCD, please see Ref. \cite{A.Bazarov.F.Karsch.S.Mukherjee.P.Petreczky.Eur.Phys.J.A55.194.2019} and references therein.

Therefore, to take into account the above discussion, one adopts the following confinement potential 
\begin{eqnarray}\label{eq:cornell_2}
V_d(r)=\sqrt{2\pi}\sigma(r-r_{\tiny{\mbox{0}}})e^{-\frac{2\pi}{\lambda^2}{(r-r_{\tiny{\mbox{0}}})^2}}.
\end{eqnarray}

Despite the arbitrary aspect of the potential (\ref{eq:cornell_2}), it preserves the confinement scale of (\ref{eq:model_2.1}) as well as gives to the hadron a finite size depending explicitly on $\lambda$. It is interesting to note that the damped confinement potential introduced by Blaschke {\it et al.} \cite{D.Blaschke.O.Kaczmarek.E.Laermann.V.Yudichev.Eur.Phys.J.C43.81.2005}, partially based on the work of Dixit \cite{V.V.Dixit.Mod.Phys.Lett.A5.227.1990} and using the running coupling constant given by the regularized method of Shirkov \cite{D.V.Shirkov.Theor.Math.Phys.136.1.893.2003}, has an asymptotic long-range behavior given by
\begin{eqnarray}\label{eq:blaschke}
V(r)\sim \frac{e^{-(r/\lambda)^2}}{r^{1/2}},
\end{eqnarray}

\noindent which differs from the long-range behavior of (\ref{eq:cornell_2}), given by 
\begin{eqnarray}\label{eq:longrange}
V_d(r)\sim re^{-(r/\lambda)^2}.
\end{eqnarray}

The factor $1/r^{1/2}$ implies the long-range (\ref{eq:blaschke}) decreases faster than (\ref{eq:longrange}) for the same physical conditions. It is important to stress here the confinement potential (\ref{eq:cornell_2}) is inspired by the Maxwell-Boltzmann distribution function. However, instead of describing the particle velocity in some homogeneous gas, one describes the static spatial separation between the $q\bar{q}$-pairs.

The damped potential (\ref{eq:cornell_2}) presents a maximum at $r^{\tiny{\mbox{max}}}$ (and a minimum, $r_{\tiny{\mbox{min}}}$), given by the simple calculation of $dV_d(r)/dr=0$
\begin{eqnarray}\label{eq:rmax}
r_{\tiny{\mbox{min}}}^{\tiny{\mbox{max}}}(\lambda)=r_{\tiny{\mbox{0}}}\pm\frac{\lambda}{2\sqrt{\pi}},
\end{eqnarray}

\noindent where $r^{\tiny{\mbox{max}}}$ corresponds to the positive sign (\ref{eq:rmax}), and one assumes the effective range of the potential (\ref{eq:cornell_2}) is restricted to $r_{\tiny{\mbox{min}}}\leq r \leq r^{\tiny{\mbox{max}}}$. Thus, the index $2\pi(r-r_{\tiny{\mbox{0}}})^2$ describes an effective area while the factor $\sqrt{2\pi(r-r_{\tiny{\mbox{0}}})}$ represents the energy contained in that area.

Notice that in the relativistic case $\lambda<\!\!<1.0$ fm, resulting in $r_{\tiny{\mbox{min}}}\approx r^{\tiny{\mbox{max}}}\approx r_{\tiny{\mbox{0}}}$. In contrast, in the non-relativistic case, it is interesting to note that from lattice QCD, the string breaking is characterized by the distance $r=1.13(10)(10)$ fm, i.e. this value represents the largest allowable distance between the constituents of the pair \cite{G.S.Bali.T.Dussel.T.Leippert.H.Neff.Z.Prkacin.KSchilling.Nucl.Phys.Proc.Suppl.153.9.2006}. Thus, as an example, one can assume in the non-relativistic case that $\lambda\approx r_p\approx r^{\tiny{\mbox{max}}}$, where $r_p$ is the proton radius. Then, one obtains
\begin{eqnarray}
r_p\approx \frac{2\sqrt{\pi}r_{\tiny{\mbox{0}}}}{2\sqrt{\pi}-1}\approx 0.83 ~\mathrm{fm}.
\end{eqnarray}
	
The above result can be compared with the recent measurement of the proton radius \cite{W.Xiong.etal.Nature.575.147.2019}, $r_p\approx 0.843(1)$ fm, indicating a very good agreement. Yet, if the one adopts $r_{\tiny{\mbox{0}}}\approx 0.47$ fm from \cite{A.Gray.etal.Phys.Rev.D72.094507.2005}, then one has $r_p\approx 0.65$ fm, a result $\approx 29\%$ lower than the experimental measurement.

In spite of the arbitrariness in the construction of the potential, the precise knowledge of $\lambda$ may be important to describe the hadron radius and the physical quantities depending on it.

\section{Total Cross Section}\label{sec:tcs}

It is well-known that in the deep-inelastic scattering, the structure of the hadron can be described by the parton distribution function, which includes the probe scale $Q$ and the Bjorken variable $x$. The so-called factorization theorems combined with perturbative techniques in QCD allows the obtaining of the hadronic scattering cross section. 

In the domain of the forward elastic scattering, the total cross section is one of the key physical quantities, being given by the sum of elastic ($\sigma_{el}(s)$) and inelastic ($\sigma_{in}(s)$) cross-sections
\begin{eqnarray}\label{eq:tot_1}
\sigma_{tot}(s)=\sigma_{el}(s)+\sigma_{in}(s).
\end{eqnarray}

\begin{figure}
\centering{	\vspace{-0.5cm}
	\includegraphics[scale=0.4]{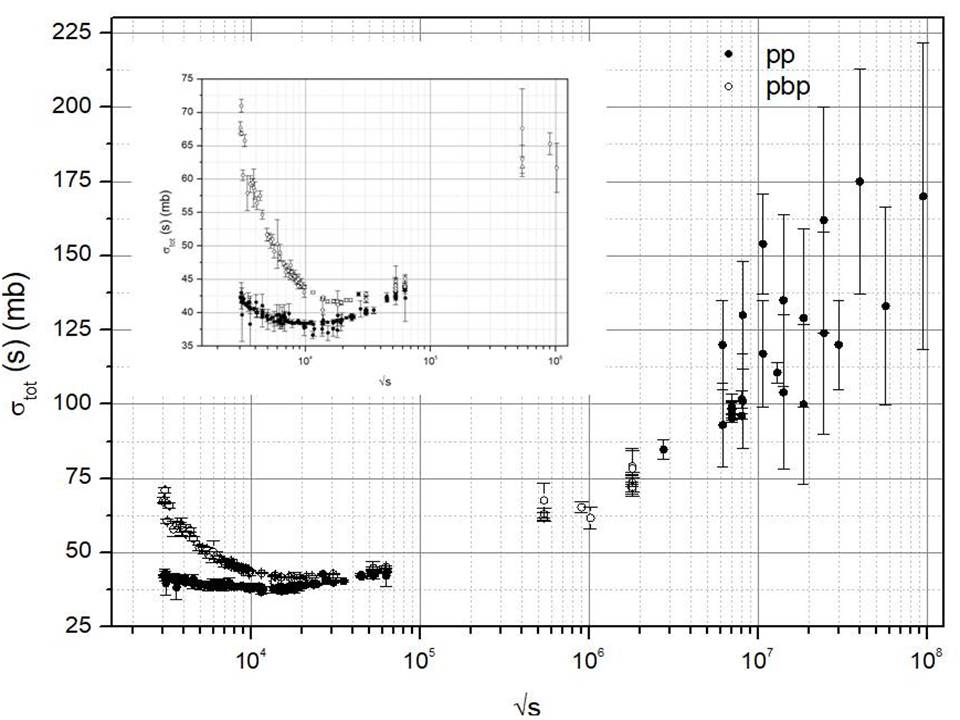}
	\vspace{-0.3cm}
	\caption{\label{fig:fig_sigma}Proton-proton and proton-antiproton ($pbp$ in the plot) total cross section experimental data
		 \cite{PDG-PhysRev-D98-030001-2018}, including the cosmic-ray results. The inner panel explicit the differences between $pp$ and $p\bar{p}$ data near the minimum of total cross section. Despite the reaction, the hadronic total cross section seems to present the same general dependence with the collision energy \cite{PDG-PhysRev-D98-030001-2018}.}}
	\end{figure}

The prediction of the rise of $\sigma_{tot}(s)$ with $s$ is a victory of massive Quantum Electrodynamics (QED) \cite{cheng_phys_rev_lett_24_1456_1970,cheng_phys_rev_d1_1064_1970}. However, nor the {\it original dynamics} of the proton (pre-rise) neither its {\it new dynamics} (post-rise) are well-understood in terms of QCD formalism. Of course, the dynamics of the $pp$ and proton-antiproton ($p\bar{p}$) elastic scattering depends on how their internal constituents can absorb and convert the incoming energy from the beam in the physical features measured in the laboratory. Figure \ref{fig:fig_sigma} shows the experimental data above $\sqrt{s}=3.0$ GeV up to (including) the cosmic-ray data for $pp$ and $p\bar{p}$ total cross section \cite{PDG-PhysRev-D98-030001-2018}. This figure summarizes the general trend presented by the hadronic total cross sections as the collision energy grows. 

In the face of the absence of general comprehension on the total cross section behavior, there are formal theoretical results from the Axiomatic Quantum Field Theory used to constrain the total cross section as the collision energy grows. One of that results is the Froissart-Martin bound used, in general, in phenomenological models to tame the growth of $\sigma_{tot}(s)$ as $s$ rise. This theoretical achievement can be written as \cite{froissart_phys_rev_123_1053_1961,martin_nuovo_cim_42_930_1966}
\begin{eqnarray}\label{eq:fm}
\sigma_{tot}(s)\leq c\ln^2(s/s_{\tiny{\mbox{0}}}),
\end{eqnarray}

\noindent where $c$ and $s_{\tiny{\mbox{0}}}$ are convenient parameters. 

Recently, a novel theoretical approach for the leading Regge pole was introduced \cite{S.D.Campos.Phys.Scr.95.6.2020,S.D.Campos.Chin.Phys.C.2020}. In that approach, the leading Regge pole can be represented by the logarithmic of the collision energy, which means this representation agrees with the Froissart-Martin bound. Based on this result, one can use the following parameterization to fit the total cross section experimental data \cite{S.D.Campos.Chin.Phys.C.2020}
\begin{eqnarray}\label{sigma_fit}
\sigma_{tot}(s)\approx \frac{a_1}{(s/s_{\tiny{\mbox{0}}})^{a_2}}+a_3\ln^{\alpha_\mathbb{P}(0)}(s/s_{\tiny{\mbox{0}}}),
\end{eqnarray}

\noindent where $a_i$, $i=1,2,3$ are free fitting parameters. The parameter $\alpha_\mathbb{P}(0)$ is the pomeron intercept, and considering only experimental data above 1.0 TeV and taking into account the recent TOTEM measurement of $\rho(s)$ \cite{G.Antchev.etal.TOTEM.Coll.Eur.Phys.J.C79.785.2019}, one has $\alpha_\mathbb{P}(0)=1.05\pm0.05$ \cite{S.D.Campos.Chin.Phys.C.2020}. 
Using the experimental data for $pp$ above 3.0 GeV up to the cosmic-ray and $\sqrt{s_{\tiny{\mbox{0}}}}=1.0$ GeV and neglecting the $\rho(s)$-parameter influence on the fitting, one obtains the parameters displayed in Table \ref{tab:table_1}. The value for $\alpha_\mathbb{P}(0)$ is typical of the hard pomeron \cite{V.S.Fadin.E.A.Kuraev.L.N.Lipatov.Sov.Phys.JETP44.443.1976,Y.Y.Balitsky.L.N.Lipatov.Sov.J.Nucl.Phys.28.822.1978}.

\begin{table*}[ht]

{\begin{tabular}{c | c | c | c | c}
			
		\hline
		 $a_1$ (mb)     ~&~ $a_2$          ~&~ $a_3$ (mb)    ~&~ $\alpha_\mathbb{P}(0)$ ~&~ $\chi^2/ndf$ \\ 
		\hline
		 $52.52\pm0.38$ ~&~  $0.14\pm0.01$ ~&~ $0.91\pm0.08$~&~ $1.61\pm0.03$ ~&~ 1.77 \\
		\hline
	\end{tabular}\caption{Parameters obtained by using (\ref{sigma_fit}) in the fitting procedure assuming  $\sqrt{s_{\tiny{\mbox{0}}}}=1.0$ GeV.\label{tab:table_1}}}

\end{table*}

The main reason to neglect the influence of recent measurement of $\rho$ in the fitting procedure, as will be seen, is because the focus of the problem is the decreasing sector of $\sigma_{tot}$, which has a small influence from $\rho$. As commented early, the total cross section is assumed here in its classical interpretation. 
Keeping in mind this assumption, the resulting total cross section can be written as
\begin{eqnarray}\label{eq:tot_2}
\sigma_{tot}(r_{h})\approx 2\pi r_{h}^2.
\end{eqnarray}

Regardless of naive, this classical interpretation allows an intuitive view of the total cross section, enabling a straight application of the preceding results.

\section{The Debye Length and the Running Coupling}\label{sec:debye}

Strictly speaking, there is no unique definition for $\lambda$ in plasma physics. For example, in thermal equilibrium, one can say that Debye length is the distance at which the potential energy from a charge perturbation is equal to the thermal energy \cite{M.B.Kallenrode.book.2004}. In contrast, for non-equilibrium thermal systems, one can define $\lambda$ as the distance at which the potential energy available in the system drops to $e^{-1}$ \cite{D.C.Montgomery.D.A.Tidman.book.1964}. The latter definition can be used to describe a system in thermal equilibrium, whereas the first one cannot be used out-of-equilibrium. 

Thus, the peculiar nature of $\lambda$ result in different definitions. For example, the electric potential describing this effect decreases by $1/e$ for each Debye length. As the temperature increases, the screening length also grows and, at high temperatures, one may write
\begin{eqnarray}\label{eq:ansatz}
\lambda_{\scriptsize{\mbox{Plasma}}}(T)=g\sqrt{\frac{T}{T'}},
\end{eqnarray}
\noindent where $g$ is a free parameter and $T'$ is some convenient scale. However, considering finite-temperature QED, the Debye length is proportional to the system temperature \cite{J.Kapusta.B.Muller.J.Rafelski.Book.2003}, being written as
\begin{eqnarray}\label{eq:qed1}
\lambda_{\scriptsize{\mbox{QED}}}(T)\propto \frac{1}{eT},
\end{eqnarray}

\noindent representing a completely different dynamics when compared to (\ref{eq:ansatz}).

\subsection{Debye Length in the UV Domain}

In general, when the energy in the system increases, the temperature also rises. Yet, explicit dependence depends on several factors. For example, in an ideal gas, the temperature is proportional to the energy $E$ available in the system while for a light quantum gas, where particle production can occur, the temperature is proportional to $E^{1/4}$ \cite{J.Kapusta.B.Muller.J.Rafelski.Book.2003}. With the absence of information about the connection of the temperature with the collision energy $s$ available in the scattering, one adopts the simple identification $s\propto T^2$. That assumption is based on the deep inelastic electron-proton scattering produced in the high parton density regime, where the Bjorken variable can be defined as $x=Q^2/s$ \cite{goncalves.v.p.braz.j.phys.34.1406.2004}. Thus, one can write 
\begin{eqnarray}\label{eq:kt}
T\propto Q, 
\end{eqnarray}

\noindent which is in accordance with the expectation at high temperatures, near the so-called saturation scale $Q_s$ \cite{D.Kharzeev.K.Tuchin.Nucl.Phys.A753.316.2005,D.Kharzeev.Nucl.Phys.A774.315.2006,J.P.Blaizot.arxiv0911.5059v1hepph.2009}. Thus, when the collision energy grows, the distance between the constituents of the pair diminishes. Then, for a sufficient high-energy, the pair undergoes spontaneous symmetry breaking at $Q_s$, as observed in SPS and BNL.

In the UV domain of QCD, the Debye length is determined from the Debye screening mass, which has been computed at one loop perturbation in $SU\!(\!N\!)$, using perturbation theory \cite{j.i.kapusta.c.gale.finite-temperarure.field.theory.principles.and.applications.cambridge.un.press.428p.2006}, possessing a non-perturbative and gauge-invariant definition: it corresponds to the largest inverse screening Debye length among all possible Euclidean correlation function involving pair of CT-odd operators in the thermal gauge field theory \cite{p.b.arnold.l.g.yaffe.phys.rev.d52.7208.1995}. Yet, to next-to-leading order, the non-perturbative methods can no longer be applied. 
Keeping only up to the second-order term in the running coupling constant and using (\ref{eq:kt}), one can write the screening length as  \cite{A.Hart.O.Philipsen.Nucl.Phys.B572.243.2000}
\begin{eqnarray}\label{eq:deb_leng_1}
	\lambda_{\scriptsize{\mbox{UV}}}(Q)=\frac{1}{\mu_{\scriptsize{\mbox{UV}}}(Q)}=\frac{1}{\left(\frac{N}{3}+\frac{n_f}{6}\right)Q\alpha_s(Q)+\left[\gamma +\frac{N}{4\pi}\ln\left(\frac{\frac{N}{3}+\frac{n_f}{6}}{\alpha_s(Q)}\right)\right]Q\alpha_s(Q)^2},
\end{eqnarray}

\noindent where $\gamma$ is a non-perturbative constant, $\mu_{\scriptsize{\mbox{UV}}}(Q)$ is the Debye screening mass in the UV domain,  and one adopts $N=2$ for a pure Yang-Mill asymptotically free theory.

At high-temperature, one can relate the Debye screening mass to the suppression of the color confinement, i.e. to the dissociation of heavy quarkonium states in QGP \cite{t.matsui.h.saltz.phys.lett.b178.416.1986}. Recently, a mass generation mechanism for particles in QED using the Debye screening has been performed, indicating a mechanism with no symmetry breaking \cite{C.A.Bonin.G.B.deGarcia.A.A.Nogueira.B.M.Pimentel.Int.J.Mod.Phys.A35.28.2050179.2020}.

The divergence of the running coupling constant at the Landau pole (a nonphysical entity) is a matter under discussion since the confinement phase does not need a divergence to occur. For example, in AdS/QCD there is no divergence in the perturbative expansion of $\alpha_s$ \cite{S.J.Brodsky.G.F.deTeramond.A.Deur.Phys.Rev.D81.096010.2010}. Therefore, one does not expect any divergence in the Debye length when the system undergoes spontaneous symmetry breaking.

\subsection{Debye Length in the IR Domain}
	
The use of the classical view for $\sigma_{tot}$ may furnish a measure of the decreasing distance between the constituents of the $q\bar{q}$-pair. From CNI up to the minimum of the total cross section, called here the decreasing sector, one has $\sigma_{tot}\sim a_1(r/r_c)^{2a_2}$, where $r_c$ is some convenient scale. Thus, it is reasonable to suppose, as a first insight, that both the distance between the constituents of the pair as well as the Debye length can be described by the same decreasing function. Observe that from the minimum of $\sigma_{tot}$ up to the hadron dissociation, called here the rising sector, the total cross section rises but the distance between the constituents still decreases. However, the function describing this decreasing, in this energy range, may be different from that one in the decreasing sector (which may cause a different decreasing rate), for example. For the sake of simplicity, one supposes the distance between the constituents of the pair is given by the same function for both sectors of $\sigma_{tot}$.

Keeping in mind the above discussion, one proposes to study the Debye length, which corresponds here to the radius of a disk containing the $q\bar{q}$-pair in the IR domain, using the two following models
\begin{eqnarray}\label{eq:chute1}
\lambda_{\scriptsize{\mbox{IR}}}(r) \approx [\bar{a}_1(r\Lambda_{\scriptsize{\mbox{QCD}}})^{2a_2}]^{1/2},
\end{eqnarray}

\noindent and
\begin{eqnarray}\label{eq:chute2}
\lambda_{\scriptsize{\mbox{IR}}}(r) \approx k[\bar{a}_1(r\Lambda_{\scriptsize{\mbox{QCD}}})^{2a_2}]^{1/4},
\end{eqnarray}

\noindent where $\bar{a}_1=a_1/2\pi$, and $k$ has unit of $\mathrm{fm}^{1/2}$. Without loss of generality, one adopts $k=1$ $\mathrm{fm}^{1/2}$, and the parameters $a_1$, $a_2$ are those from Table \ref{tab:table_1}. 

\begin{figure}
	\centering{
		\includegraphics[scale=0.4]{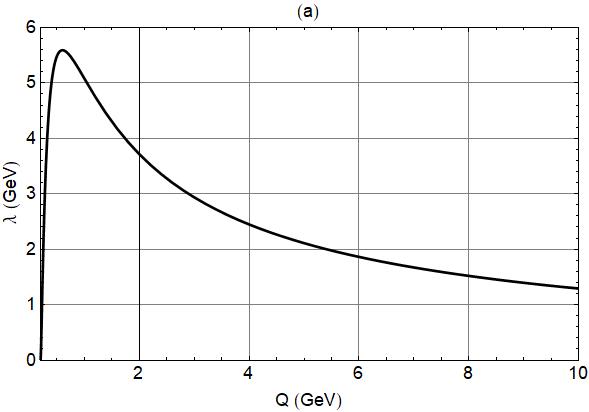}
		\includegraphics[scale=0.4]{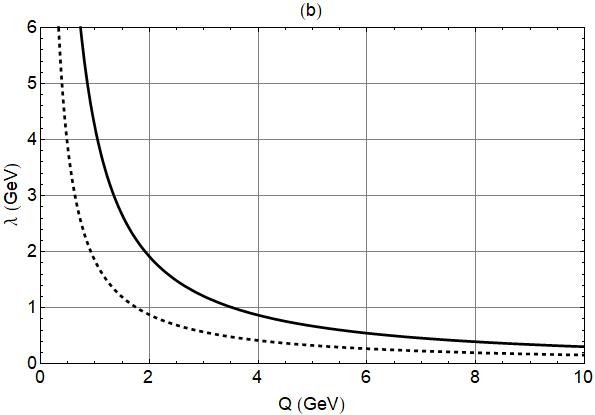} \\
		\caption{\label{fig:lambda_UV}The behavior of $\lambda$ depending on $Q$. Panel (a) shows the Debye length in non-confinement phase ($\Lambda_{\scriptsize{\mbox{QCD}}}<Q$) for (\ref{eq:deb_leng_1}) using $n_f=3$, $\gamma=1$, and $\Lambda_{\scriptsize{\mbox{QCD}}}=0.2$ GeV. In panel (b), the Fourier-sine transform of Debye length for $0<Q$: the solid line is for (\ref{eq:chute1}) and the dashed line is for (\ref{eq:chute2}).}}
\end{figure}

Figure \ref{fig:lambda_UV}(a) shows the Debye length given by (\ref{eq:deb_leng_1}) using (\ref{eq:def_alpha}) for $\Lambda_{\scriptsize{\mbox{QCD}}}<Q$. In Figure \ref{fig:lambda_UV}(b), one uses the Fourier-sine transform of $\lambda$ given by (\ref{eq:chute1}) (solid line)  and (\ref{eq:chute2}) (dashed line) considering $0< Q$. Notice that above $Q\approx 0.5$ GeV, the curves in both panels tend to vanish as $Q$ increases. 


\subsection{Running Coupling Constant in the IR Domain}

The main point here is to study the running coupling constant in the IR domain using the results of the preceding sections. To achieve more general results, one starts rewriting the confinement potential (\ref{eq:cornell_2}) as
\begin{eqnarray}\label{eq:cornell_4}
V_d(r)=\sqrt{2\pi}\sigma\left(r-2\sqrt{\frac{\alpha_s(r)}{3\sigma}}\right)e^{-\frac{2\pi}{\lambda^2}{\left(r-2\sqrt{\frac{\alpha_s(r)}{3\sigma}}\right)^2}},
\end{eqnarray}

\noindent where the root is still given by (\ref{eq:model_2.1}). From the maximum of (\ref{eq:cornell_4}), one can write the running coupling constant as
\begin{eqnarray}\label{eq:running_1}
\alpha_s(r)=\frac{3\sigma (2 \sqrt{\pi} r + \lambda)^2}{16\pi},
\end{eqnarray}

\noindent which exhibits the explicit dependence on the Debye length. To compare this result with those in literature, one use the Fourier-sine transform to write the above running coupling in $Q$-space. Replacing (\ref{eq:chute1}) and (\ref{eq:chute2}) into (\ref{eq:running_1}), one obtains, respectively,
\begin{eqnarray}\label{eq:rc_q_1}
\alpha_s(Q)=\frac{3 \sigma \cos(a_2\pi/2)\Gamma(1 + a_2) (a_1 \Lambda_{\scriptsize{\mbox{QCD}}}^{2 a_2})^{1/2}}{8\sqrt{2}\pi^{3/2}Q^{1 + a_2}}\rightarrow \alpha_s(Q)\approx \frac{1}{Q^{1+a_2}} ,
\end{eqnarray}

\noindent and

\begin{eqnarray}\label{eq:rc_q_2}
\alpha_s(Q)=\frac{3 \sigma \cos(a_2\pi/4)\Gamma(1 + a_2/2) (a_1 \Lambda_{\scriptsize{\mbox{QCD}}}^{2a_2})^{1/4}}{8\sqrt{2}\pi^{3/2}Q^{1 + a_2/2}}\rightarrow \alpha_s(Q)\approx \frac{1}{Q^{1+a_2/2}}.
\end{eqnarray}

\begin{figure}
	\centering{
		\includegraphics[scale=0.4]{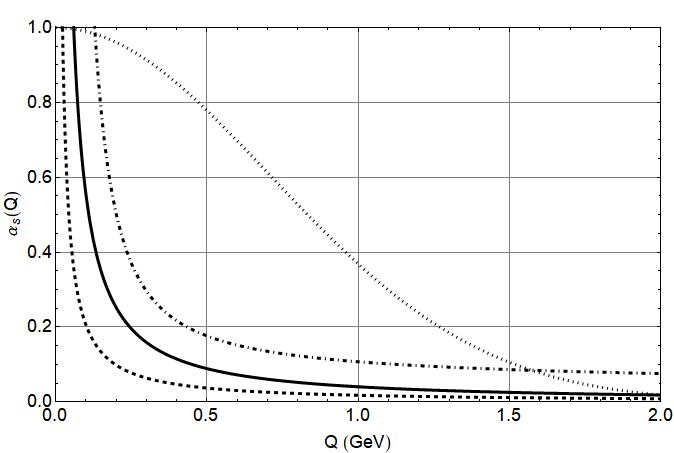}
		\caption{\label{fig:alpha_conf}The running coupling constant in the confinement phase depending on the approach for $\alpha_s(Q)$. Solid line is for (\ref{eq:rc_q_1}) and the dashed line is for (\ref{eq:rc_q_2}). The dot-dashed line is for Richardson approach (\ref{eq:richardson}) and the dotted line is for BdTD (\ref{eq:brodsky}). One uses $\kappa=0.5$ GeV, $\sqrt{\sigma}=0.4$ GeV, $n_f=3$, $N=2$, and $\Lambda_{\scriptsize{\mbox{QCD}}}=0.2$ GeV.}}
\end{figure}

As well-known, in the IR domain, the running coupling constant, obtained from perturbative techniques, cannot be used in this hadronic phase. In addition, this problem does not present a general treatment as well as there is no agreement on the definition of $\alpha_s$ in the confinement. For example, in the light-front holography context, the running coupling is given by the approach of Brodsky-de T\'eramond-Deur \cite{S.J.Brodsky.G.F.deTeramond.A.Deur.Phys.Rev.D81.096010.2010}
\begin{eqnarray}\label{eq:brodsky}
\alpha_s^{\tiny{\mbox{BdTD}}}(Q)= e^{-Q^2/4\kappa^2},
\end{eqnarray}

\noindent where one adopts here the mass scale $\kappa=0.5$ GeV. The approach due to BdTD is, of course, free of divergences.
	
Another example, from the end of the 1970s, is the method due to Richardson \cite{J.L.Richardson.Phys.Lett.B82.272.1979} to circumvent the divergence problem, defining the running coupling by adopting the prescription $(Q/\Lambda_{\scriptsize{\mbox{QCD}}})^2\rightarrow 1+(Q/\Lambda_{\scriptsize{\mbox{QCD}}})^2$ in the logarithmic term of $\alpha_s$. That prescription result in a running coupling without divergence at the Landau pole (in one-loop approximation)
\begin{eqnarray}\label{eq:richardson}
\alpha_s^{\tiny{\mbox{R}}}(Q)=\frac{1}{4\pi\beta_{0}\ln\left(1+\frac{Q^2}{\Lambda_{\scriptsize{\mbox{QCD}}}^2}\right)}.
\end{eqnarray} 

Using the running coupling given by (\ref{eq:rc_q_1}) and (\ref{eq:rc_q_2}), one can compare these results with the BdTD (\ref{eq:brodsky}) and the Richardson (\ref{eq:richardson}) approaches. The divergence-free results allow a comparison in the whole domain of $Q$ (IR and UV).

Figure \ref{fig:alpha_conf} shows the running coupling constant given by (\ref{eq:rc_q_1}) and (\ref{eq:rc_q_2}) (solid and dashed lines, respectively), $\alpha_s^{\tiny{\mbox{BdTD}}}$ (dotted line), and $\alpha_s^{\tiny{\mbox{R}}}$ (dot-dashed line) depending on $Q$. 


It is important to stress that the adoption of a unique function to describe the whole decreasing of $\lambda$ in the IR-UV domain may be a too strong assumption, which may explain the fast decreasing of $\alpha_s$ given by (\ref{eq:running_1}) (for the two $\lambda_{\scriptsize{\mbox{IR}}}$) when compared to BdTD. 

On the other hand, comparing (\ref{eq:running_1}) (for the two $\lambda_{\scriptsize{\mbox{IR}}}$) with the Richardson approach (\ref{eq:richardson}), one observes they have a close behavior, i.e. they probably has the same dynamical origin. To see that, observe that for $1\leq x_0<x$ and $0<\tau\leq 1$, the approximation 
\begin{eqnarray}
\frac{1}{(x/x_0)^{1+\tau}}\sim \frac{1}{\ln x/x_0},
\end{eqnarray}

\noindent can be used to describe the asymptotic behavior of $\alpha_s(Q)$ given by (\ref{eq:rc_q_1}) and (\ref{eq:rc_q_2}). Then, the definition (\ref{eq:def_alpha}), the one-loop approximation, can be reproduced in the asymptotic limit by the results attained here. 



It is important to note that near the saturation scale $Q_s$, the results obtained here indicates the Debye mass in the IR domain depends on $\sim T^{1+a_2}$, whereas in the UV domain $\sim T$. This difference in the behavior seems to indicate the parameter $a_2$, which has a small value in the IR domain, seems to vanish when the spontaneous symmetry breaking takes place. Thus, in the present model the phase transition occurs at the expense of $a_2$.

\subsection{Comparisons with Lattice QCD}

The effective running coupling can be defined in terms of both the spatial separation of the $q\bar{q}$-pair and the temperature by \cite{O.Kaczmarek.PoS.2008,O.Kaczmarek.F.Karsch.P.Petreczky.F.Zantow.Phys.Rev.D70.074505.2004.Erratum.ibid.D72.059903.2005} 
\begin{eqnarray}\label{eq:kaczmarek_1}
\alpha_{\tiny{\mbox{eff}}}(r,T)\equiv \frac{3}{4}r^2\frac{d}{dr}F_1(r,T)
\end{eqnarray}

\noindent where $F_1(r,T)$ is the heavy quark free energy of the static $q\bar{q}$-pair written as
\begin{eqnarray}
	F_1(\vec{r},T)=-T\log \left(\mathrm{Tr}\left(L_{ren}(\vec{0})L_{ren}(\vec{r})\right)\right),
	\end{eqnarray}

\noindent being $L_{ren}$ the renormalized Polyakov loop \cite{O.Kaczmarek.PoS.2008}. For $r\Lambda_{\scriptsize{\mbox{QCD}}}<\!<1$, the zero temperature perturbation theory allows to write \cite{O.Kaczmarek.F.Zantow.Phys.Rev.D71.114510.2005}
\begin{eqnarray}
F_1(r,T)\equiv V(r)\simeq -\frac{4}{3}\frac{\alpha_s(r)}{r},
\end{eqnarray}

\noindent while for $rT>\!>1$, with $T>T_c$, one has 
\begin{eqnarray}
F_1(r,T)\simeq -\frac{4}{3}\frac{\alpha_s(T)}{r}e^{-r/\lambda(T)}.
\end{eqnarray}	
	
One can define the entropy at asymptotic large distances, $F(r=\infty,T)$, as \cite{O.Kaczmarek.PoS.2008}
\begin{eqnarray}\label{eq:kaczmarek_2}
	S(r=\infty,T)=-\frac{\partial F(r=\infty,T)}{\partial T}.
	\end{eqnarray}

This approach is particularly interesting for $r=\infty$ since it indicates that entropy contribution is a fundamental quantity due to its fast rise near the phase transition \cite{O.Kaczmarek.PoS.2008}. That result is corroborated by \cite{S.D.Campos.A.M.Amarante.Int.J.Mod.Phys.A35.2050095.2020}, where the growing entropy near the phase transition eventually leads to the hadron dissociation.

Near the saturation scale $T\approx Q_s$, the results for the running coupling obtained here can be compared, for example, with $\alpha_{\tiny{\mbox{eff}}}(r,T)$ \cite{O.Kaczmarek.PoS.2008,O.Kaczmarek.F.Karsch.P.Petreczky.F.Zantow.Phys.Rev.D70.074505.2004.Erratum.ibid.D72.059903.2005}. Then, using (\ref{eq:rc_q_1}), (\ref{eq:rc_q_2}) and $Q_s=0.15$ GeV, one obtains $\alpha_{s}(Q/Q_s=1)=0.35(0.13)$, $\alpha_{s}(Q/Q_s=1.5)=0.22(0.09)$, and $\alpha_{s}(Q/Q_s=2)=0.16(0.06)$ whose values are systematically lower than \cite{O.Kaczmarek.PoS.2008} as well as strongly dependent on $\Lambda_{\scriptsize{\mbox{QCD}}}$. These lower values can be explained as well as controlled by the decreasing of the screening length in $Q$-space, i.e. the phenomenological process of modeling $\lambda_{\scriptsize{\mbox{IR}}}$ is crucial to determine $\alpha_{s}$.


\section{Summary and Conclusions}\label{sec:critical}

The precise knowledge of QCD properties at temperature up to a few GeV is far from a satisfactory level even in the equilibrium configuration. Of course, this is no reason to vindicate an approach without a theoretical and/or experimental basis. 

The confinement potential introduced here, despite the arbitrariness in its construction, presents two interesting results. The first one is the obtaining of an estimate of the hadron radius based on the Debye length, an experimental quantity, by assuming the Debye length in the non-relativistic regime is around the proton radius. One obtains, then, for the proton $r_p\approx 0.83$ fm, close to the recent experimental result $r_p\approx 0.843(1)$ fm \cite{W.Xiong.etal.Nature.575.147.2019}. The second one is the definition of the running coupling depending on $\lambda$.

It seems to be reasonable that, in the IR domain, the distance between the constituents of the $q\bar{q}$-pair decreases as the collision energy grows, and based on this assumption, one assumes that $\lambda$ also decreases. In particular, one proposes the decreasing of $\lambda$ is given by some power of the decreasing sector of $\sigma_{tot}$. The results for $\alpha_s$ based on $\lambda$ in the IR domain are compared with the approaches due to Brodsky-de T\'eramond-Deur and Richardson. Considering only the confinement phase and using the parameters obtained from the fitting procedure for the $pp$ total cross section, the approach proposed here seems to tend to the Richardson result, when $\lambda$ is given by (\ref{eq:chute1}). When the model is extrapolated to the non-confinement regime, this tendency still remains. 

The comparison with lattice QCD results \cite{O.Kaczmarek.F.Zantow.Phys.Rev.D71.114510.2005,O.Kaczmarek.PoS.2008,O.Kaczmarek.etal.Phys.Rev.D83.014504.2011} shows that the phenomenological approach performed here can be improved, demanding a deeper investigation which will be done elsewhere.



As a final criticism, one can also argue that to describe the proton with the quark potential, one should solve the bound-state problem. Of course, this is a non-classical approach, not treated here. It is important to stress, again, that the scenario presented in this paper is an attempt to classically understand the application limits of the potential theory, and also understand how the obtained results can be interpreted in a non-classical way using both the phenomenology of elastic scattering as well as the theoretical models available.

\section*{Acknowledgments}

SDC thanks to UFSCar for the financial support. 


\end{document}